\documentclass[final]{svjour3}                     % onecolumn (standard format)
\smartqed  % flush right qed marks, e.g. at end of proof
\usepackage{graphicx}
\usepackage{amsmath}
\usepackage{amssymb}
\usepackage{dcolumn}% Align table columns on decimal point
\usepackage{bm}% bold math
\usepackage{amsfonts}
\usepackage{hyperref}

\usepackage{mathptmx}      % use Times fonts if available on your TeX system
\begin{document}

\title{Decoherence in rf SQUID Qubits%\thanks{Grants or other notes
}


\author{Douglas A. Bennett \and Luigi Longobardi \and Vijay Patel \and Wei Chen \and Dmitri V. Averin \and James E. Lukens}


\institute{D. A. Bennett \and L. Longobardi \and  V. Patel \and W.
Chen \and D. V. Averin \and J. E. Lukens \at Department of Physics
and Astronomy, Stony Brook University, Stony Brook, New York
11794--3800\\
\email{Douglas.Bennett@sunysb.edu}}

\date{Received: \today / Accepted: \today}

\maketitle

\begin{abstract}

We report measurements of coherence times of an rf SQUID qubit using
pulsed microwaves and rapid flux pulses. The modified rf SQUID,
described by an double-well potential, has independent, \textit{in
situ}, controls for the tilt and barrier height of the potential.
The decay of coherent oscillations is dominated by the lifetime of
the excited state and low frequency flux noise and is consistent
with independent measurement of these quantities obtained by
microwave spectroscopy, resonant tunneling between fluxoid wells and
decay of the excited state. The oscillation's waveform is compared
to analytical results obtained for finite decay rates and detuning
and averaged over low frequency flux noise.

 \keywords{Decoherence \and Superconducting Qubits \and Flux Qubit \and SQUIDs}
\PACS{03.67.Lx, 85.125.Cp, 03.65.Yz}
\end{abstract}

\maketitle

\section{Introduction}

In the last few years there has been significant progress towards improved
coherence times in superconducting qubits
\cite{ither2005,steffen2006b,martinis2005,choirescu2003}. However, progress
has been slower in rf SQUID qubits then other similar devices such as phase
qubits \cite{steffen2006b,martinis2005} and persistent current qubits
\cite{plantenberg2007,choirescu2003,saito2006}. This lack of improved
coherence times is usually attributed to the larger size of rf SQUID devices
which require a large loop (i.e. 150 $\mathrm{\ \mu m}$ x 150 $\mathrm{\ \mu
m}$) to provide the necessary geometrical inductance. This paper will report
recent measurements of coherence times of a modified rf SQUID, including Rabi
oscillations, that suggest that the dominant source of decoherence in these
devices is low frequency flux noise from two level fluctuators as seen in
other superconducting qubits. We propose taking advantage of the sensitivity
of rf SQUIDs to low frequency flux noise to investigate the source of the flux
noise in the local environment.

An rf SQUID qubit in its simplest form consists of a superconducting ring of
inductance $L$ interrupted by a single Josephson junction with critical
current $I_{c}$ shunted by a capacitance $C$(see Fig.\ \ref{epotential}
inset). When a static external flux ($\Phi_{x}$) is applied to the SQUID loop,
it induces a screening current $I_{s}=-I_{c}\sin(2\pi\Phi/\Phi_{0})$ where
$\Phi_{0}\equiv\frac{h}{2e}$ is the flux quantum. The screened flux linking
the loop ($\Phi$) must satisfy, $\Phi=\Phi_{x}+LI_{s}$. The dynamics of this
device is analogous to a particle of mass C with kinetic energy $\frac{1}%
{2}C\dot{\Phi}^{2}$ moving in a potential given by the sum of the loop's
magnetic energy and the Josephson junctions coupling energy. Expressing the
fluxes in units of $\Phi_{0}$, this potential is \cite{friedman2000}
\begin{equation}
U(\phi,\phi_{x})=U_{0}\left[  2\pi^{2}(\phi-\phi_{x})^{2}-\beta_{L}\cos
(2\pi\phi)\right]  ,\label{epotential}%
\end{equation}
where $U_{0}\equiv\Phi_{0}^{2}/4\pi^{2}L$, $\beta_{L}\equiv2\pi LI_{c}%
/\Phi_{0}$. Lowercase ($\phi_{\_\_}$) is used to denote the various fluxes in
units of $\Phi_{0}$ (e.g. $\phi_{x}=\Phi_{x}/\Phi_{0}$). For $1<\beta_{L}<4.6$
and $\phi_{x}$ near $1/2$, the potential, shown in Fig.\ \ref{epotential}, has
the form of a double well with the two wells representing different fluxoid
states of the rf SQUID and having screening currents (of equal magnitude at
the symmetry point, $\phi_{x}=1/2$) circulating in opposite directions around
the loop \cite{friedman2000}. Changing $\phi_{x}$ has the effect of tilting
the double well potential. One can select either the lowest level in each well
or the two lowest levels in a single well for the basis states of the qubit.
Since the two states have an easily measurable flux difference in the double
well basis, this is referred to as the flux basis and requires a relatively
low barrier to provide adequate coupling between the states. The single well
mode, where the basis states are coupled by microwave radiation, on the other
hand, requires a relatively high barrier so that the basis states do not
couple to the states in the other well. This mode of operation is very close
the that of qubits commonly referred to as \textquotedblleft
phase\textquotedblright\ qubits, so we will use this term to identify it.
Figure \ref{figpotlev} shows an example of calculated energy levels for the
potential as a function of $\phi_{x}$. As can be seen, changes in $\phi_{x}$
cause a much greater change (about $100\times$) in the level spacing for the
flux basis than for the phase basis. Since, as we shall see, there is
substantial noise in $\phi_{x}$, this gives a strong advantage to the phase
basis in terms of coherence times.

\begin{figure}\centering\includegraphics[width=2.8in]{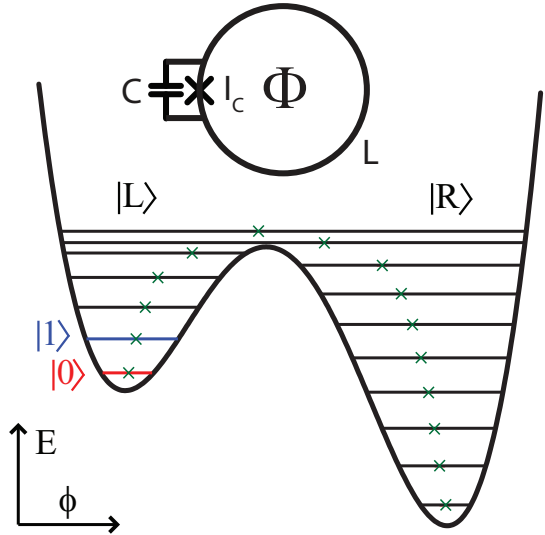}\caption{(Color
online) The potential of an rf-SQUID at a $\beta=1.32$ and
$\phi_{x}=0.505$ showing localized energy levels and the
corresponding value of mean flux
(green x). The inset shows a schematic representation of an rf SQUID}%
\label{figpotlev}%
\end{figure}

\begin{figure}\centering\includegraphics[width=3.1in]{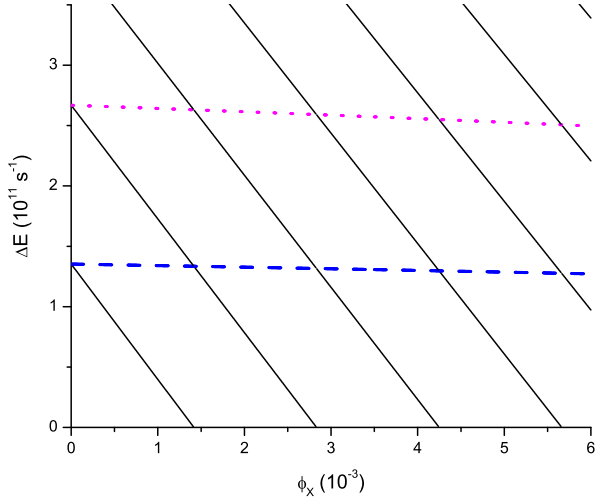}\caption{(Color
online) The energy levels of the rf SQUID referenced to the lowest
energy level in the left well. The blue dashed line (magenta dotted
line) is the first (second) excited state localized in the left
well. The black lines are
the levels localized in the right well.}%
\label{figpotlev}%
\end{figure}

\section{Design and Fabrication of the Qubit and On-Chip Control and Read-Out
Circuitry}

\label{sect:qubitchip}

\begin{figure}\centering\includegraphics[width=3.3in]{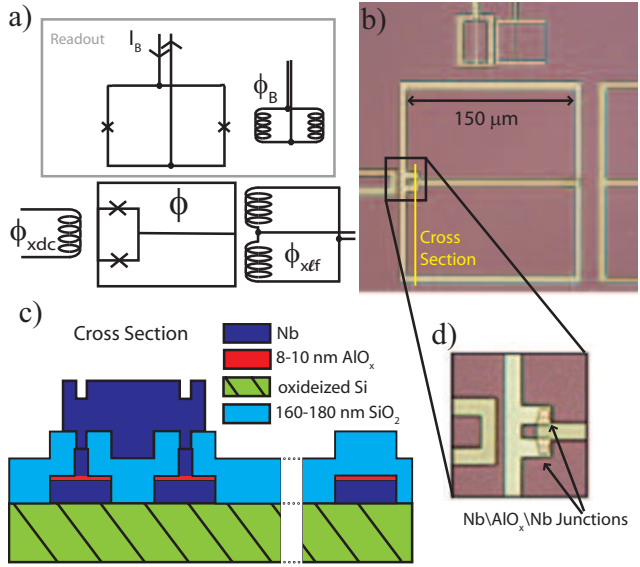}\caption{(Color
online) (a) Schematic and (b) micrograph of an rf SQUID qubit and
the readout magnetometer, (c) a cross section of the wafer around
the junctions and (d)
micrograph giving a detailed view of junctions. }%
\label{schematicfig}%
\end{figure}

The qubits used in this work, shown schematically in the bottom center of
Fig.\ \ref{schematicfig}(a), are enhanced versions of the simple qubit
discussed above. This design uses a gradiometric configuration to help isolate
the qubit from fluctuations in the ambient flux on the length scale (150
$\mathrm{\mu m}$) of the qubit, but can still be described by the simple
potential of Eq.\ \ref{epotential}. If a single flux quantum were trapped in
the outer loop when the qubit cooled through its transition temperature, then
the motion of the system between the two wells of the potential would
correspond to motion of the flux from the upper to lower loop of the
gradiometer through the junctions connecting these two loops. If the ambient
flux were zero, this situation would also result in the qubit potential being
symmetric without further bias flux.

As noted above, different modes of operation of the qubit requires changes in
the barrier height. It would, therefore, be very convenient to be able to
adjust the height of the barrier of the double well potential, \textit{in
situ}, independently of $\phi_{x}$. This requires the ability to vary the
critical current of the Josephson junction coupling the two wells. The twin
requirements of low capacitance and rapid modulation using small control
currents mean that a single large area junction cannot be used for this
purpose.\ However, the critical current can be adjusted if the single junction
is replaced by two junctions, with critical currents $I_{c1}$ and $I_{c2}$, in
a small loop, a dc-SQUID, with inductance $\ell$ as shown in
Fig.\ \ref{schematicfig}(a). By applying a flux, $\phi_{xdc}$, to this small
loop the critical current of the dc-SQUID is suppressed allowing the barrier
of the double well potential to be modulated. The addition of a second
junction formally makes the potential two dimensional. However, if $L\gg\ell$
and $I_{c1}\approx I_{c2}$, then this 2D potential is very well approximated
by the one dimensional potential in Eq.\ \ref{epotential} with small
corrections \cite{han1992,han1989} which can be neglected. $\beta_{L}$ is now
given by $\beta_{L}(\phi_{xdc})=\beta_{L0}\cos(\pi\phi_{xdc})$ where
$\beta_{L0}\equiv2\pi L(I_{c1}+I_{c2})/\Phi_{0}$.

The flux $\phi_{x}$ that tilts the potential has both a low frequency
component, $\phi_{x\ell f},$ that is applied using the on-chip gradient coil
shown at the bottom right of Fig.\ \ref{schematicfig}(a) and high frequency
components, discussed below in sec. III. The gradiometric design of the qubit
has the further advantage that, to first order, there is no cross-coupling
between the tilting flux, $\phi_{x\ell f},$ and the barrier modulation flux,
$\phi_{xdc}.$ The outer loop of the $150$ $\mathrm{\mu m}$ $\times$ $150$
$\mathrm{\mu m}$ qubit is a 5 $\mu m$ wide Nb film with a thickness of 150 nm
as is the film separating the two halves of the gradiometer except the
thickness is 200 nm. This gives an effective qubit inductance of $L\approx215$
pH. The dc-SQUID loop is $5$ $\mathrm{\mu m}$ $\times$ $5$ $\mathrm{\mu m}$
giving $\ell\approx10$ pH. These inductances are calculated using the 3D-MLSI
software package \cite{khapaev2001} and are consistent with measurements.

The fluxoid state of the qubit is measured using a hysteretic dc SQUID
magnetometer shown at the top of Fig.\ \ref{schematicfig}(a). The measured
mutual inductance between the qubit and the magnetometer is 5 pH, consistent
with the design value of 4.3 pH. The self inductance of the magnetometer is 56
pH and the junctions are 2.15 $\times$ 2.15 $\mathrm{\mu m}$ and 2.85 $\times$
2.85 $\mathrm{\mu m}$ giving a maximum critical current of $10.8$ $\mathrm{\mu
A}$. This asymmetric design, along with the asymmetry in the magnetometer flux
bias loop, $\phi_{b}$, is part of a scheme to decouple the magnetometer bias
leads from the qubit to first order while permitting fast triggering of the readout.

These circuits are fabricated at Stony Brook in our laboratory using a niobium
trilayer process called the\textquotedblleft SAL-EBL\textquotedblright%
\ process \cite{patel2005,chen2004}. This process has been developed from our
well established planarized process for Nb superconducting electronics -- the
PARTS-EBL process \cite{patel1999}. The SAL-EBL process was developed
specifically to study effects of various process steps on coherence in Nb
based qubits. In particular, to maximize junction and film
quality\cite{patel2005,pottorf08} and to minimize the turn-around time only
essential process steps were retained, eliminating CMP (chemical mechanical
polishing) and reducing process steps such as reactive ion etches (RIE) known
to reduce film quality under certain conditions\cite{chen2003}. The resulting
process has only one RIE step and uses a self-aligned lift-off (SAL) of the
dielectric to define the junctions. By using lift-off for majority of the
patterning, the process allows material flexibility for dielectric and wiring
metal, and by using a combination of deep UV and electron beam lithography
(EBL) on the same resist layer at each step allows great design flexibility.

The devices are fabricated on 2" oxidized Si wafers. The $\mathrm{Nb/AlO_{x}%
/Nb}$ trilayer is patterned via lift-off. Both Nb and Al are dc sputtered in
Ar and $\mathrm{AlO_{x}}$ is formed by thermal oxidation. The junctions
defined by EBL on a negative resist are formed by RIE in $\mathrm{SF_{6}}$ gas
with a minimum device size of 0.15 x 0.15 $\mathrm{\mu m^{2}}$. The high area
uniformity of the junctions allow us to accurately control junction asymmetry.
The same resist mask is used to lift off rf sputtered $\mathrm{SiO_{2}}$
dielectric. A Nb wiring layer is also patterned through lift-off.
Fig.\ \ref{schematicfig}(c) shows a cross-section of a completed device
indicating various layer thicknesses. The process turn-around time is
approximately one week, and the use of EBL for all critical features enables
rapid design changes.

The size for the qubit junctions vary depending on desired qubit
parameters and is 1.45 x 1.45 $\mathrm{\mu m^{2}}$ for the data
shown. The designed current density for these devices is 1
$\mathrm{\mu A/\mu m^{2}}$. Both the junction size and current
density are measured on diagnostic chips and agree with designed
values. In particular the measurements of electrical size from both
current and resistance scaling are consistent with physical size of
the junctions estimated from RIE undercut. The asymmetry of the
junctions (1.6\%) is extracted from fits to $\phi$ as function of
$\phi_{x}$ and is small enough to justify the use of Eq.\
\ref{epotential} \cite{bennettthesis}.

\section{Apparatus and Measurement Techniques}

\label{sect:apparatus}

Two essential criteria for the measurement of qubit dynamics are that one be
able to prepare and readout states of the qubit with a time resolution short
compared to the decoherence time of the states being studied and that the
qubit be very well isolated from its environment so that the decoherence time
is not affected, e.g.\ by the measurement and control processes, while the
qubit is undergoing coherent evolution. This section presents the approaches
used to achieve these goals with nanosecond resolution for qubits using the
phase basis, which is far less sensitive to low frequency noise in $\phi_{x}$.

The on-chip bias coils, discussed in Sec. \ref{sect:qubitchip}, can apply
fluxes of the order of $\Phi_{0}$ but are limited to frequencies of less than
$100$ kHz. To achieve nanosecond timing resolution, a separate set of high
frequency flux bias signals, a microwave ($\phi_{xrf}$) and a video
($\phi_{xp}$) pulse with lengths on the order of nanoseconds and
sub-nanosecond rise times$,$ are inductively coupled to the qubit via a hole
in the ground plane of a well characterized microstrip transmission line
\cite{bennett2007} on a separate chip suspended 25 $\mathrm{\mu m}$ above the
qubit chip . These signals, while very fast, are limited in amplitude to
several thousandth of $\Phi_{0}.$ This gives, for the total bias flux applied
to the large loop of the qubit, $\phi_{x}=\phi_{x\ell f}+\phi_{xrf}+\phi
_{xp}+\phi_{off},$ where $\phi_{off}$ is a constant that includes the ambient
flux and the zero offset in $\phi_{x\ell f}$, which is taken to be 0 at the
symmetry point of the potential$.$ The desired state of the qubit is obtained
from the initial ground state of the left well by applying one or more
microwave and video pulses.

Since the magnetometer cannot distinguish between the two basis states in the
phase basis, an intermediate step must be used for the readout of the qubit
state. To accomplish this, at the moment readout is desired, a short pulse,
$\phi_{xp},$\ is applied to tilt the potential rapidly but adiabatically to
reduce the tunneling barrier between wells. The barrier height during this
tilt is about $1.1$ K for the excited state of the qubit giving a tunneling
rate to the right well of about 10$^{9}$ $\mathrm{s^{-1}}$ while the rate for
the ground state is about a factor of 500 slower. Thus it is possible to
clearly distinguish these two states by appropriate selection of the pulse
length. After the end of the pulse, the system will be trapped in either the
left well (ground state) or the right well (excited state) for at least $1$ s.
Subsequent readout of the occupied well using the magnetometer will then
reveal the state of the qubit when $\phi_{xp}$\ was applied.

\begin{figure}\centering\includegraphics[width=3.2in]{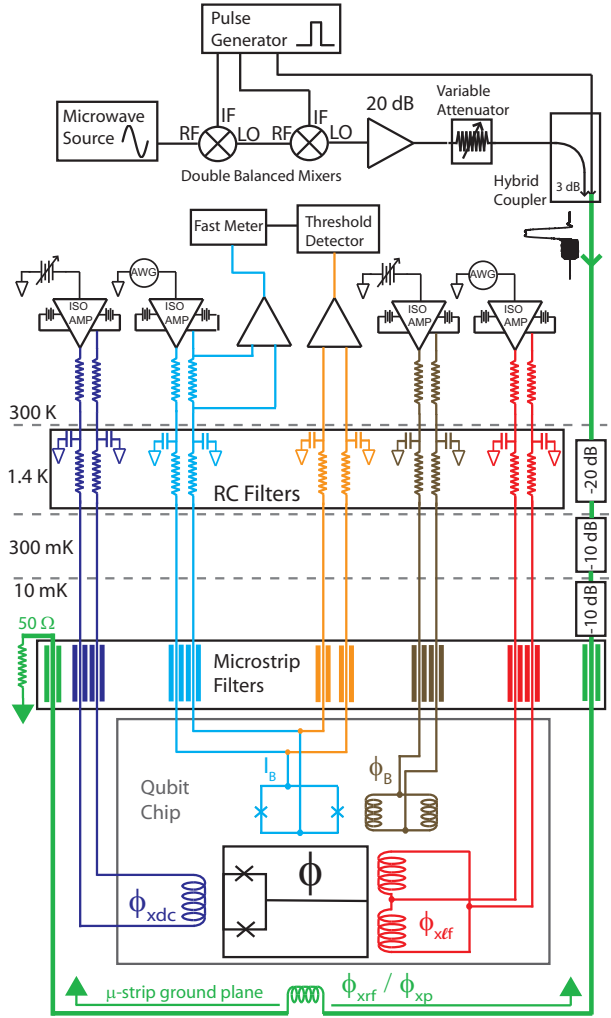}\caption{(Color
online) Circuit schematic including the various stages of filtering
and the components used to create and combine $\phi_{xrf}$ and $\phi_{xp}$.}%
\label{fig:circuitdiag}%
\end{figure}

The components that create and combine the microwave pulses and
readout pulse are shown at the top of Fig.\ \ref{fig:circuitdiag}.
The pulse generator is capable of producing measurement pulses with
rise times as short as 200 ps. Two microwave mixers are used to
modulate the envelope of the continuous microwave signal produced by
the microwave source giving an on/off ratio of $10^3$. The output of
the mixers is then amplified 20 dB and coupled through variable
attenuators that set the amplitude of the microwave pulses applied
to the qubit. Finally, these microwave pulses are combined with the
video pulses using a hybrid coupler. The video pulses enter on the
directly coupled port while the microwave pulse are coupled using
the indirectly coupled port. The video pulses used for the data
shown in the following sections are around typically 5 ns with a
rise time of 0.5 ns. The directionality and frequency response of
the coupler allows the two signals to be combined with a minimum
amount of reflections and loss of power. This signal is then coupled
to the qubit through a coaxial line that is filtered by a series of
attenuators at 1.4 K (20 dB), 600 mK (10 dB) and at the qubit
temperature (10 dB) followed by a lossy microstrip filter that cuts
off around 1 GHz.

The low frequency bias and readout circuits for the qubit and magnetometer
also have been optimized to control and measure the qubit while protecting it
from noise. All signal paths are designed to be symmetric with respect to the
chip to reduce the effect of common mode noise, and grounded signal sources
are coupled through battery powered unity gain isolation amplifiers that
effectively disconnect these signals from the earth ground. These low
frequency lines are then filtered using EMI filters on entering the cryostat
so that the dewar itself acts as an rf-shield for the cold portion of the
experiment. They are further filtered using cascaded, 4-stage\ RC filters
anchored at 1.4 K. Finally the lines are filtered using specially designed
lossy microstrip filters at the qubit temperature.

Common RC filters are limited to about 1 GHz due to the parasitic inductance
of the capacitor. The lossy microstrip filters were designed to cut off
exponentially below 1 GHz and are effective to much higher frequencies. They
consist of a thin film of chromium deposited on a glass or sapphire substrate
and diced into long narrow chips. When the chips are placed in a brass
housing, which acts as a ground plane, they form a lossy microstrip
transmission line. The filters that must pass dc current also have a Nb
meander line to pass the low frequency control currents without heating the
filter assembly. These filters should be effective to the lowest order
waveguide mode of the brass housing at 63 GHz while all other modes are
greater then 500 GHz. At 18 GHz the microstrip filters have a measured
attenuation greater than 95 dB.

The qubit chip and all the necessary wires and cables are housed in a
superconducting NbTi sample cell that acts as a magnetic shield below its
transition temperature around 10 K. This sample cell is mounted in the bottom
section of a vacuum tight filter can, which has two chambers separated by an
rf tight block housing the microstrip filters. This filter can is placed
inside a double layer Cryoperm magnetic shield on the sample stage of a
dilution refrigerator capable of reaching a loaded base temperature of 5 mK.

\section{Results and Analysis \label{sec:results}}

There are a number of different ways of characterizing the coherence times in
our flux qubit. Experimentally, the simplest method of probing the decoherence
involves measuring the lineshape of macroscopic resonant tunneling (MRT)
between fluxoid states\cite{rouse1995}. However the dynamics that determine
the tunneling process are complicated, especially when the width of the peaks
are not much smaller than the spacing between them. The measurement flux pulse
discussed in the previous section makes it possible to measure the occupancy
of the excited state and study intrawell dynamics. This along with controlled
microwave pulses makes possible a direct measurement of the lifetime of the
excited state and the observation of Rabi oscillations.

\subsection{Lifetime of the Excited State \label{seclifetime}}

The decay rate between the first excited state and the ground state in the
same well ($\Gamma_{1}$) provides an upper limit on the coherence time of the
qubit in the phase basis. $\Gamma_{1}$ can be measured directly by measuring
the occupation of the excited state, $\rho_{11}$, as a function of time. For
this measurement, the qubit is pumped to the first excited state using a long
microwave pulse followed by a measurement pulse $\phi_{xp}$ to read out the
occupation of this state as a function of the delay, $\tau_{delay}$, between
the end of the microwave pulse and the measurement pulse. A 80 ns microwave
pulse, which is much longer than $T1\equiv$ $1/\Gamma_{1},$ \ is used in order
to insure a constant initial mixture of the ground and excited state.

\begin{figure}\centering\includegraphics[width=3.4in]{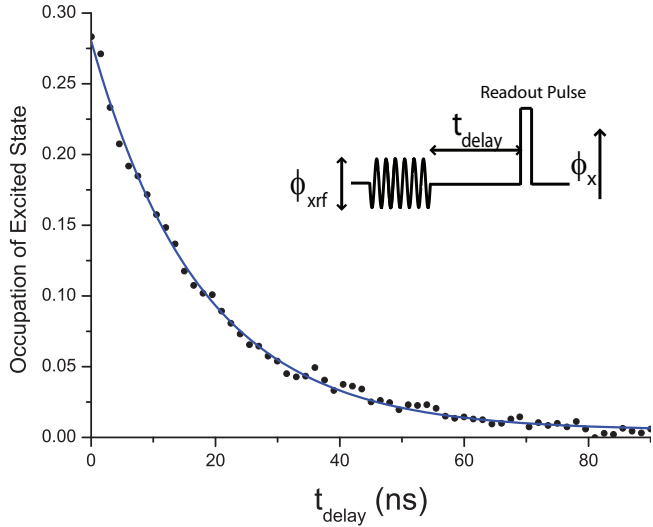}\caption{(Color
online) The measured occupation of the excited state as a function
of delay between the long microwave pulse and the readout pulse. The
line is a fit to the exponential decay used to extract $T1$. The
inset show the measurement pulse
sequence. }%
\label{decayfig}%
\end{figure}

Figure \ref{decayfig} shows the occupation of the excited state as a
function of delay between the end of the microwave pulse and the
readout pulse. The microwave frequency of 17.9 GHz is resonant with
the energy difference between the ground and excited state for the
value $\phi_{x\ell f}=0.050\times10^{-3}$ used. The solid line is
the fit to the data for $\rho_{11}\propto\exp (-\Gamma_{1}t)$\
giving $\Gamma_{1}=0.05$ $\mathrm{ns^{-1}}$ or $T1=20$ ns. As we
shall see below, such good fits are not always obtained for
arbitrary bias.

The lifetime of the excited state is similar to lifetimes measured in a number
of other $\mathrm{Nb/AlO_{x}/Nb}$ qubits operated in the phase basis (energy
eigenstates in the same potential well). Paik et al.\ have observed decay
times in dc SQUID phase qubits \cite{paik2007} around 13 ns for a
$\mathrm{Nb/AlO_{x}/Nb}$ junctions and 20 ns for $\mathrm{Al/AlO_{x}/Al}$
junctions. Martinis et al.\ observed a lifetime of 20 ns in a large area
current biased $\mathrm{Nb/AlO_{x}/Nb}$ Josephson junction \cite{martinis2002}%
. They were able to improve the lifetime to 500 ns \cite{martinis2005} through
various changes including switching to Al junctions, reducing the size of the
junction ($\sim10$ $\mathrm{\mu m^{2}}$), and changing to $\mathrm{SiN_{x}}$
for the wiring insulation layer instead of $\mathrm{SiO_{2}}$.

\subsection{Level Spectroscopy \label{secintraspec}}

\begin{figure}\centering\includegraphics[width=3.1in]{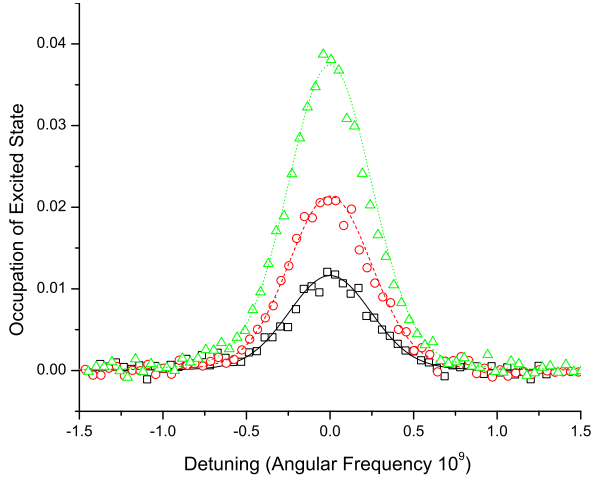}\caption{(Color
online) The occupation of the excited state as a function of
detuning for microwave powers corresponding to attenuator settings
of 39 (squares), 36 (circles), 33 dB (triangles). Lines are fits
using Eq.\ \ref{av} at microwave amplitudes corresponding to the
measured Rabi frequency for each attenuator setting (0.017, 0.024,
0.034 $\mathrm{ns^{-1}}$) with $\Gamma=0.055$ $\mathrm{ns^{-1}}$
convoluted with static Gaussian noise with $\sigma_{\nu}=0.235$
n$\mathrm{s^{-1}}$ at the angular frequencies
of the Rabi oscillations that correspond to these microwave powers}%
\label{specfig}%
\end{figure}

If the decay with lifetime $T1$ were the dominant source of decoherence in the
qubit, one would expect a resonance in the microwave occupation of the first
excited state given, in the low power limit, by $\rho_{11}\propto\phi
_{xrf}^{2}/(\delta^{2}+\Gamma_{1}^{2}/4)$ where $\delta$ is the mean
difference between the photon energy and the level spacing. For the potential
parameters used for the resonant data shown in Fig. \ref{specfig} the
dependence of $\delta$ on $\Phi_{x\ell f}$ is $\delta=1.70\times10^{9}$
$\mathrm{s^{-1}/m\Phi_{0}}$. Measurement of the equilibrium occupation of the
excited state near resonance is made using a readout pulse immediately
following a 100 ns microwave pulse, which is much longer than the lifetime of
the excited state as measured in Fig.\ \ref{decayfig}, ensuring that the qubit
reaches an equilibrium mixture of the ground and excited state. The detuning
around the resonance is accomplished by changing $\phi_{x\ell f}$, rather than
the frequency. This eliminates the effects of slight variations in $\phi
_{xrf}$\ as a function of the frequency. Figure \ref{specfig} shows the
occupation of the excited state for this potential as a function of $\delta$
for 3 different microwave power levels spaced by 3db, thus giving a factor of
two total variation in $\phi_{xrf}$. For these values, the linewidth is nearly
independent of $\phi_{xrf}$. So, one would expect the data to be in the low
power limit with a linewidth given by $\Gamma_{1}$. As can be seen, however,
the actual linewidth is about an order of magnitude greater, indicating
additional noise broadening.

The effects of extra noise on this resonance between the ground and excited
states within one well as well as on the resonant flux tunneling rate between
levels in different wells (see Sec.~\ref{MRT}) can both be described by a
similar set of evolution equations for the density matrix.

In describing both types of resonant transitions between two levels, one needs
to separate two types of interactions with the environmental noise. One is the
decay, with the rate $\Gamma,$ of the excited state to the lower-energy states
within the same well induced by the noise components at frequencies
$\simeq\omega_{p}$. Another is due to the noise components at frequencies
below $\omega_{p}$, which induce fluctuations of the energy difference
$\nu(t)$ between the resonant levels. In the case of tunneling between the
opposite wells, these fluctuations are induced directly by the energy shifts
of one well relative to the other. For transitions within one well, both
levels have practically the same value of average flux and the noise-induced
fluctuations of the energy difference $\nu$ are much smaller, roughly by two
orders of magnitude for the parameters of our qubits, than for the states in
the opposite wells. We assume that the temperature $T$ of the environment is
sufficiently large (this assumption is made more precise later in this
Section, and in Sec.~\ref{sec:rabi}), so that the fluctuations $\nu(t)$
induced by the low-frequency part of the noise can be treated classically. In
this case, evolution equations for the elements $\rho_{ij}$ of the density
matrix in the basis of the two resonant states that account for both
relaxation and the low-frequency noise can be written as (see, e.g.,
\cite{averin2000})%

\begin{align}
\dot{\rho}_{11}  &  =i\frac{a}{2}(\rho_{10}-\rho_{01})-\Gamma\rho
_{11}\,,\nonumber\\
\dot{\rho}_{00}  &  =i\frac{a}{2}(\rho_{01}-\rho_{10})+\Gamma\rho
_{11}\,,\label{rho}\\
\dot{\rho}_{01}  &  =-i\nu(t)\rho_{01}-i\frac{a}{2}(\rho_{00}-\rho_{11}%
)-\frac{\Gamma}{2}\rho_{01}\,,\nonumber\\
\dot{\rho}_{10}  &  =i\nu(t)\rho_{10}+i\frac{a}{2}(\rho_{00}-\rho_{11}%
)-\frac{\Gamma}{2}\rho_{10}\,.\nonumber
\end{align}
The coupling amplitude $a$ of the two states stands here for the tunnel
amplitude $\Delta$ in the case of flux tunneling between the two wells, or for
the microwave amplitude, $\phi_{xrf},$ expressed as the Rabi frequency
$\Omega$\ on resonance\ in the weak relaxation/dephasing limit
(Sec.~\ref{sec:rabi}).

For microwave excitation of the first excited state in the well, $\Gamma$ is
equal to the relaxation rate $\Gamma_{1}$ of this first excited state, and
$\nu(t)$ in Eqs.~(\ref{rho}) is the detuning between the excitation energy of
this state and microwave frequency. If the low-frequency noise is negligible,
$\nu(t)$ reduces to the average detuning $\delta$, and the steady-state
occupation of the excited states is obtained from the stationary solution of
the Eqs.~(\ref{rho}) and has a Lorentzian line shape as a function of $\delta
$:
\begin{equation}
\rho_{11}(\delta)=\frac{\Omega^{2}}{4\delta^{2}+2\Omega^{2}+\Gamma_{1}^{2}}.
\label{lorentzshape}%
\end{equation}
If the cut-off frequency $\omega_{c}$ of the flux noise is much lower than the
actual linewidth $W$ of the resonant transition, determined self-consistently
by the rms value $u$ of the magnitude of the low-frequency noise and by the
relaxation rate $\Gamma_{1}$ (see Eq.~(\ref{av}) below), the noise can be
accounted for simply by convoluting the intrinsic Lorentzian lineshape
(\ref{lorentzshape}) with some static distribution of the detuning values.
Under a natural assumption of the Gaussian noise, this gives
\begin{equation}
\langle\rho_{11}\rangle=\frac{1}{\sqrt{2\pi}u}\int d\nu e^{-(\nu-\delta
)^{2}/2u^{2}}\rho_{11}(\nu)\,. \label{av}%
\end{equation}

In principle, the shape of the resonant peaks can provide information not only
on the magnitude of noise but also on its spectral properties. For this, one
needs to solve Eqs.~(\ref{rho}) for the time-dependent $\nu(t)=\delta
+\tilde{\nu}(t)$. This can be done analytically for the small microwave
amplitude, $\Omega\ll W$, which can be treated as perturbation, i.e., when the
contribution of $\Omega$ to the peak broadening (\ref{lorentzshape}) is
negligible. Spectral properties of the stationary classical Gaussian noise are
characterized by the correlator:
\begin{equation}
\langle\tilde{\nu}(t_{1})\tilde{\nu}(t_{2})\rangle=\frac{1}{2\pi}\int d\omega
S(\omega)e^{i\omega(t_{1}-t_{2})}\,. \label{noise}%
\end{equation}
Perturbation theory in the coupling amplitude can be done directly in
Eqs.~(\ref{rho}) by integrating the equation for one of the off-diagonal
elements of $\rho$ under the assumption that diagonal elements are constant,
substituting then the result in the equations for the diagonal elements, and
finally averaging the resulting transition rate over the Gaussian fluctuations
of detuning using Eq.~(\ref{noise}). This straightforward calculation gives
\begin{align}
\langle\rho_{11}\rangle &  =\frac{\Omega^{2}}{2\Gamma_{1}}\int_{0}^{\infty
}d\tau e^{-\Gamma_{1}\tau/2}\cos\delta\tau\times\nonumber\\
&  \exp\{-\frac{1}{\pi}\int d\omega S(\omega)\frac{\sin^{2}\omega\tau
/2}{\omega^{2}}\}\,. \label{cutoff}%
\end{align}
This result is the limit of the classical noise, reached for large
temperatures, $T\gg W$, of a more general quantum expression
obtained in \cite{amin2008}. Qualitatively, the quantum component of
the noise leads to a finite shift of the resonant peak position away
from the zero-detuning point $\delta=0$ \cite{harris2008}. The shift
disappears with increasing temperature $T$.

For noise with a Lorentzian spectrum:
\begin{equation}
S(\omega)=\frac{2u^{2}/\omega_{c}}{1+(\omega/\omega_{c})^{2}}\,, \label{lor}%
\end{equation}
Eq.~(\ref{cutoff}) gives
\begin{align}
\langle\rho_{11}\rangle &  =\frac{\Omega^{2}}{2\Gamma_{1}}\int_{0}^{\infty
}d\tau e^{-\Gamma_{1}\tau/2}\cos\delta\tau\times\nonumber\\
&  \exp\left\{  -\frac{u^{2}}{\omega_{c}}\left(  \tau-\frac{1-e^{-\omega
_{c}\tau}}{\omega_{c}}\right)  \right\}  . \label{lor}%
\end{align}
This equation describes the transition from the regime of the broadband noise,
$\omega_{c}\gg W$, when the noise simply increases the linewidth $W$ of the
Lorentzian (without the $\Omega$ in the denominator) to $\Gamma_{1}%
+2u^{2}/\omega_{c}$, to the regime of the quasistatic noise $\omega_{c}\ll W$,
when the lineshape is given by Eq.~(\ref{av}), and for strong noise,
$u\gg\Gamma_{1}$, is:
\begin{equation}
\langle\rho_{11}\rangle=\frac{\pi\Omega^{2}}{2\Gamma_{1}}\frac{1}{\sqrt{2\pi
}u}e^{-\delta^{2}/2u^{2}}. \label{Gauss}%
\end{equation}

In the case of frequently assumed $1/f$ noise of amplitude $A$:
\begin{equation}
S(\omega) = \frac{A}{|\omega|} \, , \label{1/f}%
\end{equation}
Eq.~(\ref{cutoff}) always gives the quasistatic result (\ref{av}) (with
logarithmic accuracy) with $u^{2} \simeq(4 A/\pi) \ln(W/\omega_{l})$, where
$\omega_{l}$ is the lower cutoffs of the spectrum (\ref{1/f}) roughly given by
the time of one measurement cycle.

The lines in Fig.~\ref{specfig} are fits to the data for the
microwave-induced occupation of the excited state, using Eq.
\ref{av} assuming extra low frequency flux noise noise in $\delta$
of rms amplitude $u=\sigma_{\nu}$ to obtain the lineshape. The best
fit to the data, for all the powers shown, occurs for
$\sigma_{\nu}=0.235\pm0.01$ $\mathrm{ns}^{-1}$ and
$\Gamma_{1}=0.055\pm0.005$ $\mathrm{ns^{-1}}$. This analysis is
based on the assumption that $\phi_{x}$ remains constant during the
approximately $100$ ns time of a single measurement (the
quasi-static approximation) suggesting that low frequency flux noise
is indeed the dominant source of resonant broadening. As a test of
the degree to which this fit actually tells us something about the
frequency dependence of the noise, we refit the lowest peak in Fig.
\ref{specfig} using Eq. \ref{lor}, which assumes Lorentzian noise
with and arbitrary cutoff, $\omega_{c}$, but is valid only in the
low power limit. The results of this are that the data can be fit
almost as well for values of $\omega_{c}$ up to $2\times10^{9}$
$\mathrm{s}^{-1}$ by increasing rms noise to
$\sigma_{\nu}=0.28\pm0.01$ $ \mathrm{ns}^{-1}$. So it would seem
that the fit provides very little information on the frequency
dependence of the noise but gives a rather firm estimate of its
amplitude.

\begin{figure}\centering\includegraphics[width=4.7in]{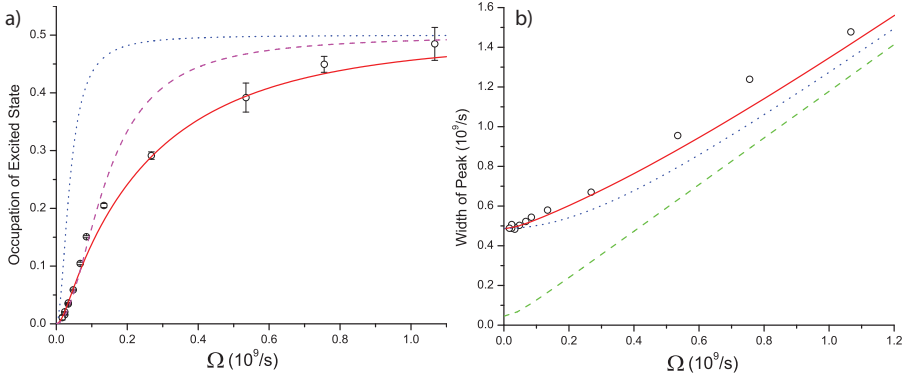}\caption{(Color
online) (a) The occupation of the excited state on resonance versus
microwave amplitude in units of the corresponding Rabi frequency.
The lines are calculations for the following parameters; red solid
$\Gamma=0.055$ $\mathrm{ns^{-1}}$ and $\sigma_{\nu}=0.235$
n$\mathrm{s^{-1}}$, blue dotted line $\sigma_{\nu}=0$ with
$\Gamma=0.055$ $\mathrm{ns^{-1}}$, purple
dashed line $\sigma_{\delta}=0$ and $\Gamma=0.20$ $\mathrm{ns^{-1}}%
$. (b) The width of the spectroscopic peak from the Gaussian fits as
a function of microwave amplitude in units of Rabi frequency. The
lines are calculations for the following parameters; red solid
$\Gamma=0.055$ $\mathrm{ns^{-1}}$ and $\sigma_{\nu}=0.235$
n$\mathrm{s^{-1}}$, green dashed line $\sigma_{\nu}=0$ with
$\Gamma=0.055$ $\mathrm{ns^{-1}}$, blue dotted line $\sigma_{\nu}=0$
and $\Gamma=0.585$ $\mathrm{ns^{-1}}$.}
\label{specamp}%
\end{figure}

\bigskip These fits can be extended to higher rf amplitudes where power
broadening is important using Eq. \ref{av}, as shown in Fig.
\ref{specamp}(a), which shows the maximum occupation of the excited
state as a function of microwave amplitude for a much wider range of
$\phi_{xrf}$. The red solid line is a fit to the peak amplitudes
($\delta=0$) using the same fitting parameters as in Fig.
\ref{specfig}. For comparison, the blue dotted line is calculated
for no flux noise, i.e. $\sigma_{\nu}=0,$ and $\Gamma=0.055$
$\mathrm{ns}^{-1}$ while the purple dashed line is calculated for
$\sigma_{\nu}=0$ and $\Gamma=0.20$ $\mathrm{ns}^{-1}$ which gives
the best fit to the low power data without considering low frequency
noise. The peak amplitude cannot be fit over the whole range of
microwave amplitudes without including the low frequency noise.
Figure \ref{specamp}(b) shows the width of the peaks as a function
of $\phi_{xrf}$ for the same data and the red solid line the fit for
the same parameters as above. The green dashed line is the width for
$\Gamma=0.055$ $\mathrm{ns}^{-1}$ and the blue dotted line for
$\Gamma=0.585$ $\mathrm{ns}^{-1}$ and no low frequency noise. Again,
low frequency Gaussian noise is required in order to obtain a
reasonable fit to the overall lineshape of the data.

\begin{figure}\centering\includegraphics[width=3.5in]{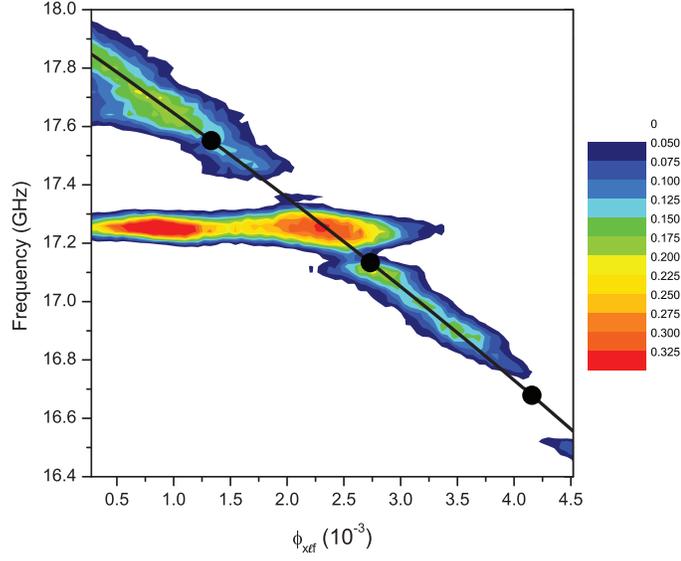}\caption{(Color
online) The measured occupation of the excited state after a long
microwave pulse expressed as color contours (blue being lowest and
red being the highest) as function of both frequency and $\phi_{x}$.
The solid lines are calculations of the energy level spitting
between consecutive eigenstates states in the same
well for $\beta=1.30$, $L=190$ pH and $C=210$ fF. }%
\label{spec2dfig}%
\end{figure}

As noted above, the data shown in Figs.\ \ref{specfig} and
\ref{specamp} were obtained by changing $\phi_{x\ell f}$\ over a
narrow range near points where the spectrum was well behaved. In
general, however, the spectrum is much more complex as can be seen
in Fig.\ \ref{spec2dfig} showing the occupation of the excited state
as function of both frequency and $\phi_{x\ell f}$ for
$\beta_{L}=1.30$ for a much wider range of parameters. The color
contours are proportional to occupation of the excited state, blue
being lower occupation and red being the highest occupation (0.325).
The curving ridge running diagonally from the top left to the bottom
right corresponds the resonance between the ground and excited
state. The black solid line shows the calculated difference between
these states as a function of $\phi_{x}$, where  $L$ and $C$ are
fitting parameters consistent with simulations and measurements (not
shown) of thermal escape rates, positions of resonant tunneling
peaks and photon assisted tunneling peaks \cite{bennettthesis}. The
gaps in the excited state occupation along this line are consistent
with coupling to two level fluctuators, e.g. in the dielectrics,
that have also been observed by other groups investigating
superconducting phase qubits
\cite{simmonds2004,martinis2005,palomaki2007}. In general these gaps
do not correspond\ to the bias values where the states in the left
well align with excited states in the right (indicated by the black
circles).

Another prominent feature in these data is the strong resonance near $17.25$
GHz over a wide range of $\phi_{x \ell f},$ which we attribute to a cavity
resonance in the NbTi cell. The larger peak of this cavity resonance, to the
left of the resonance between the ground and excited state, corresponds to a
two photon resonance to the second excited state of the qubit. The frequency
of this cavity resonance is in the range of the lowest order cavity modes of
the sample cell, even though most of these modes should be suppressed due to
the extremely small dimensions ($\sim1$ mm) of the cell out of the plane of
the chips.

\begin{figure}\centering\includegraphics[width=4.7in]{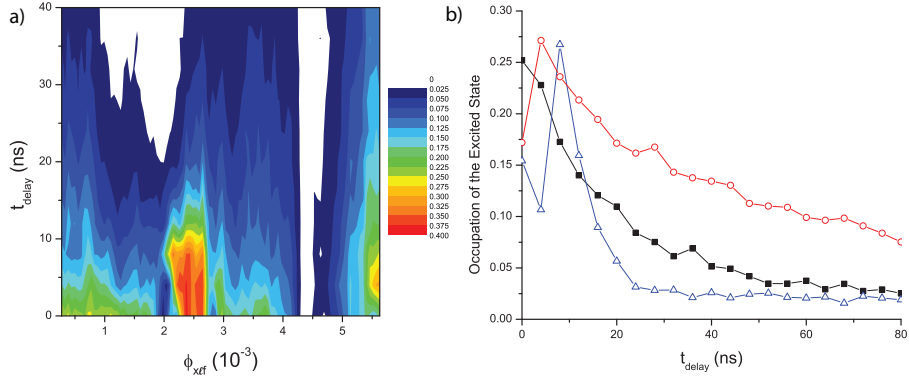}\caption{a)
The measured occupation of the excited state after a long microwave
pulse expressed as color contours (blue being lowest and red being
the highest) as function $\phi_{x}$ and the delay between the end of
the microwave pulse and the beginning of the readout pulse. b)
Slices of the contour plot at $\phi_{x\ell f}=$ $0.767\times10^{-3}$
(black squares), $2.078\times10^{-3}$ (blue triangles) and
$5.464\times10^{-3}$ (red circles). The lines are just to
help guide the eye.}%
\label{decay2dfig}%
\end{figure}

These anomalous behaviors are further illustrated in Fig.\ \ref{decay2dfig}%
(a), which shows the occupation of the excited state, using color contours, as
a function of $t_{delay}$ and $\phi_{x\ell f}$. Here, the microwave frequency
has been adjusted for each value of $\phi_{x\ell f}$ in order to remain on
resonance. The bias value used for the measurement of the lifetime of the
excited state shown in Fig.\ \ref{decayfig} is at the extreme left side of
this plot.

At a number of bias points, the decay of the excited states is
non-exponential, in contrast to the data of Fig.\ \ref{decayfig}. For example,
in the region near the gaps in the excitation spectrum, there are actually
peaks in the occupation of the excited state as a function of $t_{delay}$.
Figure \ref{decay2dfig}(b) shows slices of Fig.\ \ref{decay2dfig}(a) at
$\phi_{x\ell f}=$ $0.767\times10^{-3}$ (black squares), $2.078\times10^{-3}$
(blue triangles) and $5.464\times10^{-3}$ (red circles). At $\phi_{x\ell f}=$
$0.767\times10^{-3}$ the decay is similar to that of Fig.\ \ref{decayfig},
while for the other two $\phi_{x}$ values the behavior is clearly not
exponential. The occupation of the excited state actually increases with time
and reaches a local maximum at a non zero delay. At these bias points, it
appears that the qubit is coupled strongly to a two level fluctuator as was
reported by Cooper et\ al.\ \cite{cooper2004} in their investigation on
current-biased phase qubits. They also observed an oscillation superimposed on
the decay of the excited state near one of these splittings in the excitation spectrum.

\subsection{Comparison with Resonant Tunneling Peaks\label{MRT}}

As noted above the same equations (with different parameters) that
determine the shape of the spectroscopic peaks discussed above
should also describe the resonance tunneling between levels in
different wells of the potential as initially seen in experiments on
MRT, where an enhancement of the escape rate from one well to the
other was observed when the lowest energy level of the initial well
approximately aligns with an energy level in the other. When the
energy difference between the levels ($\varepsilon$) in opposite
wells is much less than $\omega_{p}$ and the intrawell energy
relaxation from the final state (the $n^{\text{th}}$ level of the
right well) is $\Gamma_{n}$, this tunneling rate, which has a form
analogous to Eq. \ref{lorentzshape}, is given by \cite{averin2000}
\begin{equation}
\Gamma_{esc}=\frac{\Delta^{2}\Gamma_{n}}{2\Delta^{2}+\Gamma_{n}^{2}%
+4\varepsilon^{2}}. \label{lorentzpeak}%
\end{equation}

Away from resonance when the levels are localized in their respective wells,
the escape rate can be approximated as the sum of the tunneling rates from the
ground state in the left well ($|i\rangle$) to the lower energy states in the
right well ($|f\rangle$). Assuming Ohmic dampening, this gives for the
individual rates \cite{kopietz1988}
\begin{equation}
\gamma_{_{i},_{f}}=\frac{2\pi}{\hbar}\frac{R_{Q}}{R}\left(  E_{i}%
-E_{f}\right)  \left\vert \langle i\left\vert \bm{\mathrm{\Phi}}
\right\vert f\rangle\right\vert ^{2},\text{ and
}\Gamma_{esc}=\sum\gamma_{_{i},_{f}}
\label{equ:tunnelij}%
\end{equation}
where R is the ohmic dampening, $R_{Q}\equiv h/4e^{2}$ is the
quantum unit of resistance and $\bm{\mathrm{\Phi}}$ is the flux
operator.

\begin{figure}
\begin{center}
\includegraphics[width=3.0in]{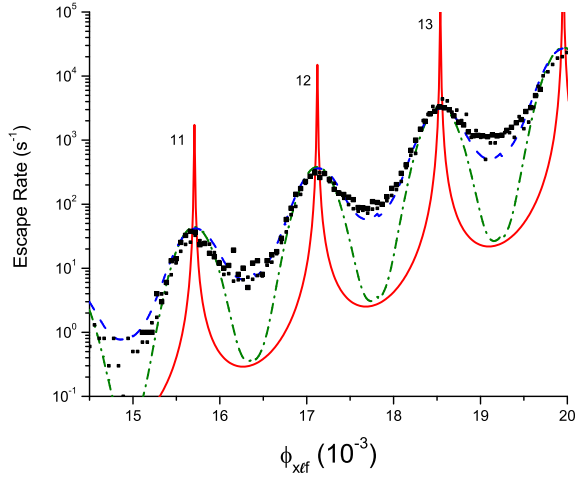} \caption{(Color online)
Measured escape rates from the ground state in the left well as a
function of $\phi_{x}$ at a $\beta=1.412$. The solid red line is a
calculation for $\Gamma_{12}=0.68\operatorname{ns}^{-1}$. The green
dash dot line corresponds to the solid line with the addition of
static Gaussian noise with
$\sigma_{\nu}=0.23\operatorname{ns}^{-1}$. The blue dashed line has
the same level of Gaussian noise but with $\Gamma_{12}=10.5\operatorname{ns}%
^{-1}$. }%
\label{ratesfitfig}%
\end{center}
\end{figure}

These rates can be measured by pulsing $\phi_{x}$ from some initial
value,\ where $\Gamma_{esc}$\ in negligible to the tilt at which $\Gamma
_{esc}$ is to be measured and leaving it there for some time $t_{m}$, in which
the system has some reasonable probability of tunneling though the barrier
into an excited state of the other flux well. Under the assumption above of
incoherent tunneling, the system will then decay to the ground state of the
new fluxoid well and remain trapped, so many repetitions of this measurement
will give the probability of tunneling. The escape rate is determined from
this probability of a transition, $P(\phi_{x})$, by
\begin{equation}
\Gamma_{esc}(\phi_{x})=\frac{1}{t_{m}}\ln\left(  \frac{1}{1-P(\phi_{x}%
)}\right)  .
\end{equation}

The measured escape rates are shown in Fig.\ \ref{ratesfitfig} as a
function of $\phi_{x\ell f}$. The resonant tunneling peaks are
labeled with an index that indicates which level in the right well
is aligned with the ground state in the left well. The red solid
line is a calculation of the escape rates using Eqs.
\ref{lorentzpeak} and \ref{equ:tunnelij} with the measured qubit
parameters. The decay rate, $\Gamma_{12}=0.68$ $\mathrm{ns}^{-1}$ is
the product of the decay rate directly measured in Section
\ref{seclifetime} and the level number ($n=12$). This calculation
gives peaks much narrower than the measured peaks. The green dash
dot lines are these same peaks convoluted with static Gaussian flux
noise in analogy to Eq. \ref{av} with $\sigma_{\nu}=0.235\pm0.01$ $
\mathrm{ns}^{-1}$, the same as the level of flux noise used to fit
the spectroscopy in Sec.\ \ref{secintraspec}. For these parameters,
the shape of the resonant tunneling peaks are well fit, but not the
intervening valleys. In order to make both the peaks and valleys
fit, the relaxation rate must be increased to $\Gamma_{12}=10.5$
$\mathrm{ns}^{-1}$ which is inconsistent with the measured
$\Gamma_{1}$. It may well be that a more refined theory is needed
for the tunneling minima between the peaks.

\subsection{Time domain measurements \label{sec:rabi}}

While the data in the previous sections, clearly show the effects of various
processes leading to decoherence in the qubit and even provide a means to
obtain qualitative estimates of the decoherence rates, they are all obtained
from non-coherent processes. In order to demonstrate coherence in our qubit
and to compare the resulting decoherence rates with those due to the know
sources of decoherence discussed above, we now turn to time domain
measurements of the evolution of coherent superpositions of the two qubit states.

An analytic solution to the Bloch equations for the population of the excited
state in the presence of CW microwave, assuming no decoherence and that the
system starts in the ground state at time zero, is given by%

\begin{equation}
\rho_{11}(t,\delta)=\frac{\Omega^{2}}{\Omega^{2}+\delta^{2}}\sin^{2}%
(\Omega_{Rabi}(\delta)t/2), \label{equ:rabi}%
\end{equation}
where
\begin{equation}
\Omega_{Rabi}(\delta)=\sqrt{\delta^{2}+\Omega^{2}}. \label{rabifreq}%
\end{equation}

This solution, shows that when driven with microwaves, the {population} of the
excited states oscillates in time, demonstrating the phenomenon of Rabi
oscillations, which have been seen in a number of different superconducting
qubits \cite{nakamura1999,choirescu2003,vion2002,martinis2002,wallraff2005}.
$\Omega,$\ the frequency of the oscillations for $\delta=0,$ is ideally
proportional to the amplitude of the microwaves excitation $\phi_{xrf}$ . This
linear dependence is accurately seen in our qubit at low power levels (see
Fig. \ref{rabifig} upper inset) providing a convenient means to calibrate the
amplitude of the microwaves incident on the qubit in terms of $\Omega.$

The measurement sequence for coherent oscillations is very similar
to that used to measure the lifetime of the excited state except
that the duration of the (generally shorter) microwave pulse is
varied and the microwave pulse is immediately followed by the
readout flux pulse. This signal, on the high frequency line, is
illustrated in the lower inset in Figure \ref{rabifig}. Figure
\ref{rabifig} shows an example of the Rabi oscillations when
$\delta=0$ and the microwave frequency, $f_{xrf}=17.9 \mathrm{GHz}$.
This bias point lies in a "clean" range of the spectrum ($17.6-18
\mathrm{GHz})$ as seen in Fig. \ref{spec2dfig}, which should be the
best region for observing coherent oscillations between the ground
and excited states. Most of the time domain data, including the
lifetime measurements of Fig.\ \ref{decayfig}, have been taken in
this frequency range. Each data point in Fig. \ref{spec2dfig}
corresponds to the average of several thousand measurements for a
given pulse length.

\subsection{Rabi Oscillations \label{sec:rabi_theory}}

\bigskip

In order to simulate the time dependent dynamics of the Rabi
oscillations in the presence of the low-frequency noise, it is
convenient to separate this noise in the two parts. One, the
quasi-static part, consists on the noise components with frequencies
smaller than the linewidth $W$ of the corresponding resonance. Under
the conditions of the Rabi oscillations, $W$ is smaller than the
strength $\Omega$ of coherent rf coupling between the two levels,
which gives the Rabi oscillation frequency in the case of resonance
($\delta=0$) and weak relaxation/dephasing. This part of the noise
can be taken into account by simple averaging over the static
distribution of detuning $\delta$ as in Eq.~\ref{av}. The second,
``dynamic'' part of the low-frequency noise has components with
frequencies extending beyond $W$. We treat this noise making an
assumption (relevant for the data presented in this work) that the
temperature $T$ is larger than not only $W$, but also the Rabi
frequency $\Omega$. In this case, it is possible to characterize
this part of the noise as the classical noise with the same constant
spectral density $S$ extending above $\Omega$ and equal to the
spectral density at frequency $\Omega$. As discussed in
Sec.~\ref{secintraspec}, such a broadband noise simply increase the
decoherence rate of the two states by $S/2$ (i.e., by
$u^{2}/\omega_{c}$ in the example of Eq.~(\ref{lor}) of that
Section) to $\gamma= (\Gamma+S)/2$. With this modification of the
decoherence rate, Eqs.~(\ref{rho}) for the evolution of the density
matrix of the two rf-coupled states can be written as
\begin{align}
\dot{p}  &  =\Omega v-\Gamma p-\Gamma/2\,,\nonumber\\
\dot{u}  &  =\delta v-\gamma u\,,\;\;\;\;\;\;\label{Bl}\\
\dot{v}  &  =-\delta u-\Omega p-\gamma v\, ,\nonumber
\end{align}
in terms of the imbalance $p=(\rho_{11}-\rho_{00})/2$ of the occupation
probabilities $\rho_{00}, \, \rho_{11}$ of the ground and excited state, and
the real/imaginary parts $u,v$ of the off-diagonal element $\rho_{01}=u+iv$ of
the density matrix. As Bloch equations (\ref{Bl}) describe the dynamics of the
Rabi oscillations, and are appropriate when the relaxation/dephasing rates are
small on the scale of plasma frequency. The finite relaxation/dephasing rates
make the analytic solution to these equations, which is shown below,
considerably more complex than the undamped solution of Eq.~\ref{equ:rabi}.

Decaying Rabi oscillations described by Eqs.~(\ref{Bl}) lead to some
stationary values of the occupation imbalance $p$ and coherences $u,v$ that
are established after the decay of the oscillations. The analytic solution of
Eqs.~(\ref{Bl}) that corresponds to this process can be written as:
\begin{equation}
A(t)=A_{st}+e^{-\gamma t}\sum_{j}e^{\lambda_{j}t}A_{j}\,,
\label{sol}
\end{equation}
where the \textquotedblleft vectors\textquotedblright\ $A$ are composed of the
elements (\ref{Bl}) of the density matrix, $A=\{p,u,v\}$, and $A_{st}$
represents the stationary part of the solution:
\begin{equation}
A_{st}=\frac{-1/2}{\delta^{2}+\gamma^{2}+\Omega^{2}\gamma/\Gamma}\{\delta
^{2}+\gamma^{2},\Omega\gamma,\Omega\delta\}\,.\label{st}%
\end{equation}
The sum in Eq.~(\ref{sol}) is taken over the three eigenvalues $\lambda_{j}$
of the evolution matrix $S$:
\begin{equation}
S=\left(
\begin{array}[c]{ccc}
\gamma-\Gamma & 0 & \Omega\\
0 & 0 & \delta\\
-\Omega & -\delta & 0
\end{array}
\right)  , \label{ev}
\end{equation}
which are given by the roots of the cubic equation
$\mbox{det}(S-\lambda)=0$. In the range of parameters relevant for
the present discussion (not very strong relaxation/dephasing) one
root is real:
\begin{equation}
\lambda_{1}\equiv\lambda=s_{+}-s_{-}+(\gamma-\Gamma)/3\,,
\end{equation}
and the other two are complex conjugated:
\begin{equation}
\lambda_{2}=\lambda_{3}^{\ast}\equiv\kappa+i\omega\,,\;\;\;\omega=\frac
{\sqrt{3}}{2}(s_{+}+s_{-})\,,
\end{equation}
\begin{equation}
\kappa=\frac{1}{2}(s_{-}-s_{+})+\frac{1}{3}(\gamma-\Gamma)\,,
\end{equation}
where
\begin{equation}
s_{\pm}=\left[  (q^{3}+r^{2})^{1/2}\pm r\right]  ^{1/3},
\end{equation}
and
\begin{align}
q &  =\frac{1}{3}(\Omega^{2}+\delta^{2})-\frac{1}{9}(\Gamma-\gamma
)^{2},\nonumber\\
r &  =\frac{\Gamma-\gamma}{3}\left(
\frac{1}{2}\Omega^{2}-\delta^{2}-\frac
{1}{9}(\Gamma-\gamma)^{2}\right)
\end{align}

The vectors $A_{j}$ in Eq.~(\ref{sol}) are the projections of the
vector $A(0)-A_{st}$ of the deviations of the initial elements of
the density matrix from their stationary values, onto the
eigenvectors of the evolution matrix (\ref{ev}) that correspond to
the eigenvalues $\lambda_{j}$. These projections can be found
conveniently from the following expression:
\begin{equation}
A_{j} = \frac{(S-\lambda_{k}) (S-\lambda_{l})}{ (\lambda_{j}- \lambda
_{k})(\lambda_{j}-\lambda_{l})} [A(0)-A_{st}] \, , \label{pr}%
\end{equation}
which for each $j$, directly cancels out components of the two other
eigenvector $k,l\neq j$ in the initial vector $A(0)-A_{st}$.

If the dynamics of the system starts at $t=0$ in the ground state, then
$A(0)=\{-1/2,0,0\}$. In this case, using Eq.~(\ref{st}) for the stationary
vector $A_{st}$ and the matrix $S$ (\ref{ev}) in the Eq.~(\ref{pr}), together
with the expressions for the eigenvalues $\lambda_{j}$ given above, we find
the time dependence of the occupation probability $\rho_{11}$ of the excited
state as
\begin{align}
\rho_{11}  &  =\frac{\gamma}{2\Gamma}\frac{\Omega^{2}}{\delta^{2}+\gamma
^{2}+\Omega^{2}\gamma/\Gamma}\left[  1-\frac{e^{-\gamma t}}{(\lambda
-\kappa)^{2}+\omega^{2}}\right. \nonumber\\
&  \Big\{e^{\lambda t}(\kappa^{2}+\omega^{2}-2\kappa\gamma+\gamma
(\gamma-\Gamma)-\Omega^{2}-\delta^{2}\Gamma/\gamma)+\nonumber\\
&  e^{\kappa t}\big[(\lambda(\lambda-2\kappa)+2\kappa\gamma-\gamma
(\gamma-\Gamma)+\Omega^{2}+\delta^{2}\Gamma/\gamma)\times\nonumber\\
&  \cos\omega t+[\omega^{2}(\gamma-\lambda)+(\kappa-\lambda)(\kappa
\lambda-\gamma(\kappa+\lambda)+\nonumber\\
&  \left.  \gamma(\gamma-\Gamma)-\Omega^{2}-\delta^{2}\Gamma/\gamma
)]\frac{\sin\omega t}{\omega}\big]\Big \}\right]  . \label{fin}%
\end{align}
Equation (\ref{fin}) gives the decay rate of the Rabi oscillations as
$\eta=\gamma-\kappa$. If the decay/relaxation rates in Eqs.~(\ref{Bl}) are
small in comparison to $\Omega$, the rate $\eta$ reduces to the Rabi
oscillations decay rate $\widetilde{\Gamma2}$ obtained previously by Ithier et
al. \cite{ither2005}, who expressed it in terms of the noise-induced decay
rate at the Rabi frequency $\Gamma_{\nu}$ and the decoherence rate
$\Gamma_{\phi}$ as
\begin{equation}
\widetilde{\Gamma2}=\frac{1}{\delta^{2}+\Omega^{2}}\left[  \frac{\Gamma}%
{4}\left(  3\Omega^{2}+2\delta^{2}\right)  +\delta^{2}\Gamma_{\phi}%
+\frac{\Gamma_{\nu}\Omega^{2}}{2}\right]  \,. \label{equ:rabidecay}%
\end{equation}
As discussed above, at large temperature $T$, the noise is classical at the
Rabi frequency, and one can take $\Gamma_{\nu}=\Gamma_{\phi}$. In this case,
$\gamma=\Gamma/2+\Gamma_{\nu}$, and for weak relaxation/dephasing, the two
Rabi decay rates agree, $\eta(\delta)=\widetilde{\Gamma2}(\delta)$.

\begin{figure}\centering
\includegraphics[width=3.3in]{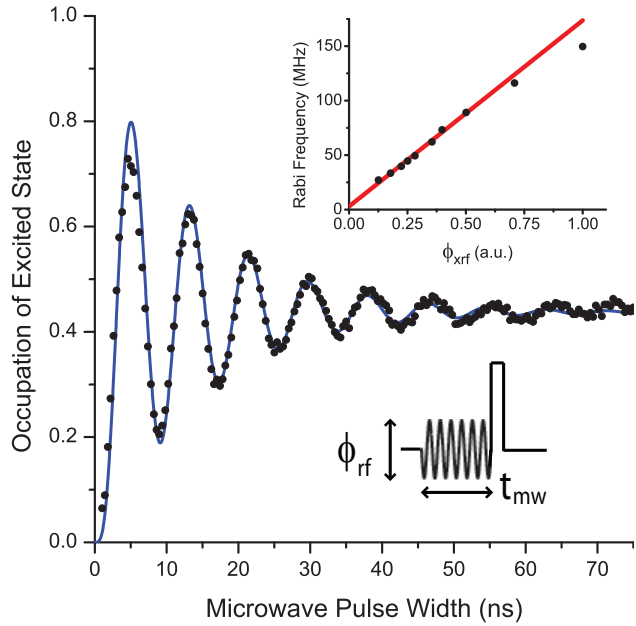} \caption{(Color
online) The occupation of the excited state as a function of the
length of the microwave pulse demonstrating Rabi oscillations. The
line is a fit to Eq. \ref{fin} for $\delta=0$ averaged over
quasi-static noise with $\sigma_{\nu }=0.22$ $\mathrm{ns}^{-1}$.
This gives $f_{rabi}=119$ MHz and decay time $\widetilde {T2}=24$
$\mathrm{ns}^{-1}$. The upper inset shows Rabi frequency as a
function of amplitude of applied microwaves in arbitrary units. The
line is a linear fit to the lower microwave amplitude data. The
lower inset show the measurement pulse sequence.} \label{rabifig}
\end{figure}

The data for resonant Rabi oscillations, shown in Fig.
\ref{rabifig}. The solid line is a fit\ using Eq. \ref{fin}
including a $0.5$ ns time delay for the rise time for the microwave
pulse and averaged over quasi-static Gaussian noise in $\phi_{x}$
equal to that obtained from the fits to the spectroscopy data in
Fig. \ref{specfig}. This gives a Rabi frequency $f_{rabi}=119
\mathrm{MHz}$ and decay rate, $\eta=0.042$ $\mathrm{ns}^{-1}$, for
the oscillations. Equation \ref{equ:rabidecay}, with $\delta=0$,
together with our previously measured value for the decay rate of
the first excited state $\Gamma_{1}$ (which gives the rate $\Gamma$
in the equations above), and the observed Rabi decay rate $\eta$
discussed below, imply $\Gamma_{\nu}= 0.01$ $\mathrm{ns}^{-1}.$\
Even though $\Omega_{Rabi}$ is not affected by flux noise to first
order for $\delta=0,$ the amplitude of the flux noise in our qubit
is large enough to cause a measurable effect. If the flux noise were
neglected, it would be necessary to increase $\eta$ to $0.058$
$\mathrm{ns}^{-1}$ in order to account for the observed decay. In
addition to increasing the decay rate, the low frequency noise also
reduces the steady state occupation of the excited state$,$ i.e. its
value for long pulses, as calculated from Eq. \ref{av}. The ordinate
of Fig. \ref{specfig} has been calibrated using this value and is
consistent with the estimate obtained \ from the calculated
tunneling rates from the two levels involved during the readout
pulse.

The upper inset to Figure \ref{rabifig} shows the Rabi frequency
extracted from Rabi oscillation measurements versus microwave
amplitude in arbitrary units. As noted above, at low powers the Rabi
frequency is linear in microwave amplitude as expected. At higher
powers, however, it begins to saturate. There is evidence that at
these higher power levels there is a small probability of excitation
to the second excited state. However since the second excited state
has orders of magnitude higher escape rate, this small probability
can be significant.

\begin{figure}\centering\includegraphics[width=3.3in]{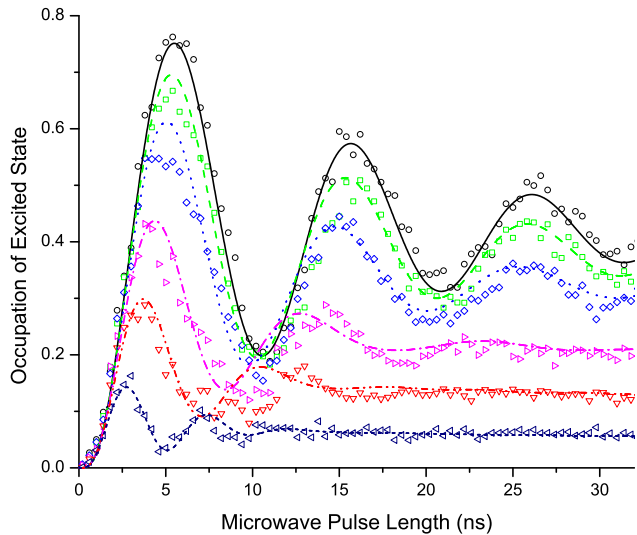}\caption{(Color
online) Rabi oscillations for detunings going from top to bottom of
0.094, 0.211, 0.328, 0.562, 0.796 and 1.269 $\mathrm{ns^{-1}}$ with
the corresponding fits using $\Gamma=0.075$ $\mathrm{ns^{-1}}$ and
$\sigma_{\nu}=0.22\operatorname{ns}^{-1}$ }%
\label{rabidetunefig}%
\end{figure}

The effects of flux noise on the decay of the Rabi oscillation, of
course, becomes much more pronounced for $\delta\neq0$. Figure
\ref{rabidetunefig} shows such Rabi oscillations at various
detunings over a range of $0\leq \delta\lesssim10^{9}$
$\mathrm{s}^{-1} $. At long microwave pulse times (not shown) the
occupation of the excited state reaches the equilibrium values
discussed in Sec \ref{secintraspec}. Near resonance, the value of
$\rho_{11}$ for long pulse lengths is determined mainly by the noise
amplitude, while for large $\delta$ it is set by $\delta.$ The data
cannot be fit to solutions of the Bloch equations using only the
phenomenological decay constants. As in the spectroscopy, static
Gaussian detuning noise must be included to fit the data over the
whole range of detuning. The lines in Figure \ref{rabidetunefig}
correspond to calculations with an initial Rabi frequency of $0.59$
n$\mathrm{s^{-1}}$, $\Gamma =0.075$ $\mathrm{ns^{-1}}$ and
$\sigma_{\nu}=0.22$ $\mathrm{ns^{-1}}$. The detuning frequency is
taken from $\phi_{x}$ using the conversion from Fig.\
\ref{spec2dfig}. For the range of $\delta$ in Fig.\
\ref{rabidetunefig} $\Delta E_{01}$\ is in a region with relatively
few spectral anomalies. However for $\delta<0$, on the other side of
resonance towards splittings in the spectroscopy, the data generally
do not agree with theory suggesting that the spectroscopic
splittings have a strong effect on the coherence, as expected.

\subsection{Ramsey Pulse Sequence}

\begin{figure}\centering\includegraphics[width=4.7in]{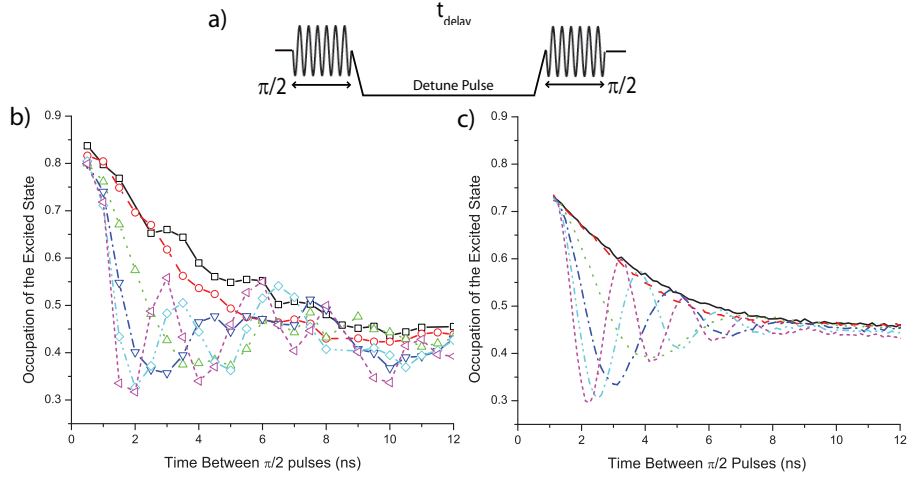}\caption{(Color
online) (a) Measurement pulse sequence and (b) Measured occupation
of the excited state after a series of 3 ns microwave pulses as a
function of delay between the two pulses for various detuning pulse
amplitudes (corresponding frequency); 0 V $\square$'s (no detuning),
0.1 V $\circ$'s (0.21 $\mathrm{ns^{-1}}$), 0.45 V $\vartriangle$'s
(0.96 $\mathrm{ns^{-1}}$), 0.8 V $\triangledown$'s (1.70
$\mathrm{ns^{-1}}$), 1.15 V $\diamond$'s (2.45 $\mathrm{ns^{-1}}$)
and 1.50 V $\triangleleft$'s (3.20 $\mathrm{ns^{-1}}$). (c)
Calculations including noise for the pulse sequence by numerically
solving the
Bloch equations for the parameters corresponding to the measurement. }%
\label{ramseyplot}%
\end{figure}

The effects of dephasing due low frequency flux noise should be seen
most dramatically in the decay of Ramsey fringes. However the
observation of Ramsey fringes in this system is complicated by the
short lifetime of the excited state. To reduce these problems we
have used the pulse sequence shown in Fig.\ \ref{ramseyplot}(a).
Here the $\pi/2$ pulses are applied on resonance and the detuning is
produced in between these pulses, using an additional $\phi_{xp}$
pulse to tilt the potential and slightly change the level spacing
for a variable time.

Figure \ref{ramseyplot}(b) shows the measured occupation of the excited state
as a function of delay between 3 ns microwave pulses (17.9 GHz) for several
different values of detuning. The lines are provided to help guide the eye
between the points. The expected Ramsey fringes can be calculated by
numerically solving the Bloch equations, using a time dependent detuning along
with the time dependent driving amplitude. As in previous sections, the
resulting oscillations are convoluted with quasi-static Gaussian distributed
noise. Figure \ref{ramseyplot}(c) shows these calculated occupations of the
excited state for the parameters corresponding to the data in
Fig.\ \ref{ramseyplot}(b). The scale factor for the amplitude of the detuning
pulses was used as a fitting parameter but is consistent with the calibration
of the detuning pulses from the ratio of the voltages between the measurement
flux pulse and the detuning pulse. The rise times for the detuning pulse was
assumed to be the same as was found for the measurement pulse. The rms value
of the low frequency noise, $\sigma_{\nu}$, observed in Sec.
\ref{secintraspec} is used for this calculation, and the decay rate
$\Gamma_{1}$ that gave the best fit to the data is consistent with that
measured in Sec. \ref{seclifetime}. The determination of $\Gamma_{1}$ has a
relatively large uncertainty due to uncertainties in the exact shape and
length of the $\pi/2$ pulses, since the short coherence times are pushing the
limits of our apparatus. However, the calculation shows qualitative agreement
with the data. The limitations of the measurement do not allow for meaningful
extraction of other parameters such as $\Gamma_{\phi}.$

\section{Sources of decoherence}

There is now a strong consensus that much of the decoherence observed in
superconducting qubits, including the flux noise seen here, comes from two
level fluctuators (TLF) in dielectrics and oxides. The evidence seems to point
towards something in the local environment as the source of the decoherence.
This is corroborated by observations of flux noise of unknown origin in SQUIDs
\cite{foglietti1986,tesche1985,wellstood1987} and persistent current qubits
\cite{yoshihara2006}.

The noise is typically independent of temperature $T$ at low
temperatures $T\leq0.1$ K. This temperature independence of the
noise, combined with the fact that the noise is produced by an
equilibrium source \cite{harris2008}, indicates that the magnetic
susceptibility of this source has paramagnetic $1/T$ dependence on
temperature \cite{send,harris2008}. While this observation supports
the theoretical models \cite{koch,rds,fi} of the flux noise produced
by weakly interacting electron spins located on the defects of the
structure, these models differ in terms of the spin precise
location. One possibility is that the spins are trapped at the
interface between the superconducting films and the insulating
substrate \cite{fi}. There is evidence
\cite{simmonds2004,cooper2004,martinis2005} that some of the common
insulating materials used for qubit fabrication have defects which
affect the coherence times of superconducting qubits. The most
likely candidates in our qubits are the substrate and the insulation
under the wiring layer. We briefly review a number of tests
conducted on our qubit to eliminate many other possible sources of
decoherence. This leaves us with the conclusion that such TLFs are
strong candidates for the decoherence seen here as well.

One extraneous possibility is that the extensive filtering of external leads
in not adequate. To check this, superconducting aluminum shunts were added to
the flux bias coils allowing the bias flux to be trapped in a superconducting
loop that included the shunt. This reduced the spectral density of any noise
coupled to the qubit by the bias lines, as well as any damping due to the
lines by two orders of magnitude but had no effect of the decoherence in the
qubit\cite{longobardi2007}. The magnetometer was maintained at zero voltage
until the qubit readout. By varying this zero voltage bias point the small
coupling between the magnetometer bias lead and the qubit could be varied by
several orders of magnitude and adjusted to zero (at least to first order).
Again, no effect on coherence was seen from this variation. Coherence was also
not affected by the repetition rate of the readouts, eliminating the
possibility that quasiparticle were being generated in the qubit by the
readouts and affecting subsequent measurements.

Back action from the magnetometer is another potential source of
decoherence. However, the magnetometer is operated in the switching
current branch which is ``silent'' until it is time to perform the
measurement. When the measurement is performed a voltage is
established across the junctions creating quasiparticles which
destroy the state of the qubit. However, for the technique used to
measure the lifetime and the Rabi oscillations, the magnetometer is
left in the superconducting state until well after the state of the
qubit was fixed. After the measurement, the magnetometer's bias
current is ramped back to zero and left there long enough that the
quasiparticles disperse. The magnetometer is designed such that its
bias current does not couple strongly to the qubit. The flux and
current biases can be set at a point where the derivative of
circulating current with respect to the bias current goes to zero.
At this point, fluctuations in the bias current do not be couple to
the qubit to first order. However, there is no difference between
operating at the optimum bias point and away from the optimum bias
point in the measurements of the lifetime of the excited state. This
implies that the magnetometer is not the major source of
decoherence.

We have also tested the quality of the junctions and Nb film used to
fabricate our qubit. Patel et al.\ \cite{patel2005} have found that
the junctions co-fabricated with our qubit devices have a subgap
resistance of 1 $\mathrm{G}\Omega$ or greater at 400 mK. This level
of damping would give coherence times orders of magnitude longer
than the values measured in Sec. \ref{sec:results}. Critical current
fluctuations, having a 1/f spectrum, have the effect of modulating
the barrier height of the rf SQUID potential resulting in
fluctuations of the energy spacing between coupled states causing
decoherence. Recent measurements by Pottorf et al.\
\cite{pottorf2007} of the level of 1/f critical current fluctuations
in junctions co-fabricated with our qubit devices have found this
noise spectrum density to be orders of magnitude less than the
commonly accepted value. Within the model of Van Harlingen et al.\
\cite{vanharlingen2004}, these critical current fluctuations are
orders of magnitude too small to explain the level of decoherence
measured in our sample.

The Nb thin film used for the roughly 0.5 mm flux loop in the qubit
could potentially have excess loss despite being in the
superconducting state. The losses at high frequencies (10's of GHz)
were tested by measuring the Qs of coplanar waveguide
resonators\cite{chen2008}. The process for fabricating these
resonators is similar to the process used to fabricate the qubit
samples. Despite discovering losses at these frequencies that seem
to depend on the fabrication process, the measured Qs ($\approx
5\times10^{5}$) were sufficiently high to rule out these losses in
the Nb films as a significant source of decoherence. We should note,
however, that these resonator measurement were at too high a power
level to exclude losses due to easily saturated TLFs. The effect of
the substrate on decoherence has been studied by fabricating and
measuring qubits and resonators on substrates of supposedly higher
quality than the 20 $\Omega \mathrm{cm}$ silicon substrates commonly
used for our qubits. These included sapphire and 15 $k \Omega
\mathrm{cm}$ silicon wafers, which showed much wider MRT peaks than
those seen in Fig.\ \ref{ratesfitfig} for qubits fabricated on a 20
$\Omega \mathrm{cm}$ silicon substrate. Further investigation is
necessary to check if the increased width is caused by the substrate
and not by some unforeseen effect on the fabrication process.

The $\mathrm{SiO_{2}}$ layer between the junction layer and wiring layer is
also a potential source of the observed decoherence. Most superconducting
qubits with longer decoherence times are made using $\mathrm{Al/AlO_{x}/Al}$
junctions \cite{bartet2005,ither2005} using a two angle shadow evaporation
process that has a thin layer $\mathrm{AlO_{x}}$ for insulation. The NIST
group uses a process the includes a deposited insulation layer. When they
switched from $\mathrm{SiO_{2}}$ to $\mathrm{SiN_{x}}$ \cite{martinis2005}
they observed a significant increase in the decay time for their Rabi oscillations.

\section{Conclusion}

With careful design of the measurement setup and a suitable fabrication
process, it is possible to observe coherent oscillations between two energy
levels in an rf SQUID. Since the quantum coherence is extremely sensitive to
external noise, extensive steps were taken to protect the rf SQUID. The sample
was fabricated using a flexible, well characterized fabrication process
capable of producing high quality Josephson junctions. The flux readout, a
hysteretic dc SQUID magnetometer, was optimized to minimize the back action on
the rf SQUID. Also the filtering and electronics were carefully designed to
reduce the amount of external noise reaching the rf SQUID. Finally a setup for
very fast pulsing of the bias flux and coupled microwaves was instituted to
allow measurement of the coherent oscillations on a nanosecond time scale.

The coherence times of the superposition between the ground and first excited
state in the same flux state were measured and analyzed within the context of
the known noise sources and calculations based on the parameters of the rf
SQUID. There exists low frequency flux noise (0.14 $\mathrm{m \Phi_{0}}$) and
a short lifetime for the excited state (20 ns) that are not consistent with
known noise sources. The coherence times are currently too short to be useful
for quantum computation. However, it is likely that the decoherence is caused
by issues relating to materials or fabrication and could be improved through a
careful comparative study.

\begin{acknowledgements}
This work was supported in part by NSF and by AFOSR and NSA through
a DURINT program.
\end{acknowledgements}

%Produces the bibliography via BibTeX.
\bibliographystyle{spphys}       % APS-like style for physics
\bibliography{decoherence}   % name your BibTeX data base

\begin{thebibliography}{10}
\providecommand{\url}[1]{{#1}}
\providecommand{\urlprefix}{URL }
\expandafter\ifx\csname urlstyle\endcsname\relax
  \providecommand{\doi}[1]{DOI \discretionary{}{}{}#1}\else
  \providecommand{\doi}{DOI \discretionary{}{}{}\begingroup
  \urlstyle{rm}\Url}\fi

\bibitem{ither2005}
G.~Ithier, E.~Collin, P.~Joyez, P.J. Meeson, D.~Vion, D.~Esteve, F.~Chiarello,
  A.~Shnirman, Y.~Makhlin, J.~Schriefl, G.~Sch\"{o}n, Physical Review B
  (Condensed Matter and Materials Physics) \textbf{72}(13), 134519 (2005).
\newblock \doi{10.1103/PhysRevB.72.134519}.
\newblock \urlprefix\url{http://link.aps.org/abstract/PRB/v72/e134519}

\bibitem{steffen2006b}
M.~Steffen, M.~Ansmann, R.C. Bialczak, N.~Katz, E.~Lucero, R.~McDermott,
  M.~Neeley, E.M. Weig, A.N. Cleland, J.M. Martinis, Science
  \textbf{313}(5792), 1423 (2006).
\newblock \doi{10.1126/science.1130886}.
\newblock
  \urlprefix\url{http://www.sciencemag.org/cgi/content/abstract/313/5792/1423}

\bibitem{martinis2005}
J.M. Martinis, K.B. Cooper, R.~McDermott, M.~Steffen, M.~Ansmann, K.D. Osborn,
  K.~Cicak, S.~Oh, D.P. Pappas, R.W. Simmonds, C.C. Yu, Physical Review Letters
  \textbf{95}(21), 210503 (2005).
\newblock \doi{10.1103/PhysRevLett.95.210503}.
\newblock \urlprefix\url{http://link.aps.org/abstract/PRL/v95/e210503}

\bibitem{choirescu2003}
I.~Chiorescu, Y.~Nakamura, C.~Harmans, J.~Mooij, Science \textbf{299}, 5614
  (2003)

\bibitem{plantenberg2007}
J.H. Plantenberg, P.C. de~Groot1, C.J.P.M. Harmans1, J.E. Mooij, Nature
  \textbf{447}, 836 (2007)

\bibitem{saito2006}
S.~Saito, T.~Meno, M.~Ueda, H.~Tanaka, K.~Semba, H.~Takayanagi, Physical Review
  Letters \textbf{96}(10), 107001 (2006).
\newblock \doi{10.1103/PhysRevLett.96.107001}.
\newblock \urlprefix\url{http://link.aps.org/abstract/PRL/v96/e107001}

\bibitem{friedman2000}
J.R. Friedman, V.~Patel, W.~Chen, S.K. Tolpygo, J.E. Lukens, Nature
  \textbf{406}, 43 (2000)

\bibitem{han1992}
S.~Han, J.~Lapointe, J.E. Lukens, Phys. Rev. B \textbf{46}(10), 6338 (1992).
\newblock \doi{10.1103/PhysRevB.46.6338}

\bibitem{han1989}
S.~Han, J.~Lapointe, J.E. Lukens, Phys.\ Rev. Lett. \textbf{63}(16), 1712
  (1989).
\newblock \doi{10.1103/PhysRevLett.63.1712}

\bibitem{khapaev2001}
M.~Khapaev, A.~Kidiyarova-Shevchenko, P.~Magnelind, M.~Kupriyanov, in
  \emph{{IEEE} Transactions on Apllied Superconductivity} (2001), pp.
  1090--1093

\bibitem{patel2005}
V.~Patel, W.~Chen, S.~Pottorf, J.E. Lukens, IEEE Transactions on Apllied
  Superconductivity \textbf{15}, 117 (2005)

\bibitem{chen2004}
W.~Chen, V.~Patel, J.E. Lukens, Microelectronic Engineering \textbf{73-74}, 767
  (2004)

\bibitem{patel1999}
V.~Patel, J.~Lukens, {IEEE} Transactions Applied Superconductivity \textbf{9},
  3247 (1999)

\bibitem{pottorf08}
S.~Pottorf, V.~Patel, J.E. Lukens, arXiv:0809.3272v1 [cond-mat.supr-con]
  (2008)

\bibitem{chen2003}
W.~Chen, V.~Patel, S.K. Tolpygo, D.~Yohannes, S.~Pottorf, J.E. Lukens, {IEEE}
  Transactions on Applied Superconductivity \textbf{13}, 103 (2003)

\bibitem{bennettthesis}
D.A. Bennett, Studies of decoherence in rf {SQUIDS}.
\newblock {PhD} dissertation, Stony Brook University, Department of Physics and
  Astronomy (2007)

\bibitem{bennett2007}
D.A. Bennett, L.~Longobardi, V.~Patel, W.~Chen, J.E. Lukens, Superconductor
  Science and Technology \textbf{20}(11), S445 (2007)

\bibitem{rouse1995}
R.~Rouse, S.~Han, J.E. Lukens, Phys. Rev. Lett. \textbf{75}(8), 1614 (1995).
\newblock \doi{10.1103/PhysRevLett.75.1614}

\bibitem{paik2007}
H.~Paik, B.K. Cooper, S.K. Dutta, R.M. Lewis, R.C. Ramos, T.A. Palomaki, A.J.
  Przybysz, A.J. Dragt, J.R. Anderson, C.J. Lobb, F.C. Wellstood, {IEEE}
  Transactions on Applied Superconductivity \textbf{17}(2), 120 (June 2007).
\newblock \doi{10.1109/TASC.2007.898124}

\bibitem{martinis2002}
J.M. Martinis, S.~Nam, J.~Aumentado, C.~Urbina, Phys. Rev. Lett.
  \textbf{89}(11), 117901 (2002).
\newblock \doi{10.1103/PhysRevLett.89.117901}

\bibitem{averin2000}
D.V. Averin, J.R. Friedman, J.E. Lukens, Phys. Rev. B \textbf{62}(17), 11802
  (2000).
\newblock \doi{10.1103/PhysRevB.62.11802}

\bibitem{amin2008}
M.H.S. Amin, D.V. Averin, Phys. Rev. Lett. \textbf{{100}}({19}) ({2008}).
\newblock \doi{{10.1103/PhysRevLett.100.197001}}

\bibitem{harris2008}
R.~Harris, M.W. Johnson, S.~Han, A.J. Berkley, J.~Johansson, P.~Bunyk,
  E.~Ladizinsky, S.~Govorkov, M.C. Thom, S.~Uchaikin, B.~Bumble, A.~Fung,
  A.~Kaul, A.~Kleinsasser, M.H.S. Amin, D.V. Averin, Phys. Rev. Lett.
  \textbf{{101}}({11}) ({2008}).
\newblock \doi{{10.1103/PhysRevLett.101.117003}}

\bibitem{simmonds2004}
R.W. Simmonds, K.M. Lang, D.A. Hite, S.~Nam, D.P. Pappas, J.M. Martinis,
  Physical Review Letters \textbf{93}(7), 077003 (2004).
\newblock \doi{10.1103/PhysRevLett.93.077003}.
\newblock \urlprefix\url{http://link.aps.org/abstract/PRL/v93/e077003}

\bibitem{palomaki2007}
T.~Palomaki, S.K. Dutta, R.M. Lewis, A.J. Przybysz, H.~Paik, B.K. Cooper,
  H.~Kwon, E.~Tiesinga, A.J. Dragt, J.R. Anerson, C.J. Lobb, F.C. Wellstood, in
  \emph{Extended Abstracts of the 11th International Superconductive
  Electronics Conference} (2007)

\bibitem{cooper2004}
K.B. Cooper, M.~Steffen, R.~McDermott, R.W. Simmonds, S.~Oh, D.A. Hite, D.P.
  Pappas, J.M. Martinis, Physical Review Letters \textbf{93}(18), 180401
  (2004).
\newblock \doi{10.1103/PhysRevLett.93.180401}.
\newblock \urlprefix\url{http://link.aps.org/abstract/PRL/v93/e180401}

\bibitem{kopietz1988}
P.~Kopietz, S.~Chakravarty, Phys. Rev. B \textbf{38}(1), 97 (1988).
\newblock \doi{10.1103/PhysRevB.38.97}

\bibitem{nakamura1999}
Y.~Nakamura, Y.~Pashkin, J.~Tsai, Nature \textbf{398}, 6730 (1999)

\bibitem{vion2002}
D.~Vion, A.~Aassime, A.~Cottet, P.~Joyez, H.~Pothier, C.~Urbina, D.~Esteve,
  M.~Devoret, Science \textbf{296}, 886 (2002)

\bibitem{wallraff2005}
A.~Wallraff, D.I. Schuster, A.~Blais, L.~Frunzio, J.~Majer, M.H. Devoret, S.M.
  Girvin, R.J. Schoelkopf, Physical Review Letters \textbf{95}(6), 060501
  (2005).
\newblock \doi{10.1103/PhysRevLett.95.060501}.
\newblock \urlprefix\url{http://link.aps.org/abstract/PRL/v95/e060501}

\bibitem{foglietti1986}
V.~Foglieitti, W.~Gallagher, M.~Ketchen, A.~Kleinsasser, R.~Koch, S.~Raider,
  R.~Sandstrom, Applied Physics Letters \textbf{49}, 1393 (1986)

\bibitem{tesche1985}
C.~Tesche, K.~Brown, A.~Callegari, M.~Chen, J.~Greiner, H.~Jones, M.~Ketchen,
  K.~Kim, A.~Kleinsasser, H.~Notarys, G.~Proto, R.~Wang, T.~Yogi, IEEE
  Transactions on Magnetics \textbf{21}, 1032 (1985)

\bibitem{wellstood1987}
F.~Wellstood, C.~Urbina, J.~Clarke, Applied Physics Letters \textbf{50}, 772
  (1987)

\bibitem{yoshihara2006}
F.~Yoshihara, K.~Harrabi, A.O. Niskanen, Y.~Nakamura, J.S. Tsai.
\newblock Decoherence of flux qubits due to 1/f flux noise (2006)

\bibitem{send}
S.~Sendelbach, D.~Hover, A.~Kittel, M.~M{\"u}ck, J.~Martinis, , R.~McDermott,
  Phys. Rev. Lett. \textbf{100}, 227006 (2008)

\bibitem{koch}
R.~Koch, D.~DiVincenzo, J.~Clarke, Phys. Rev. Lett. \textbf{98}, 267003 (2007)

\bibitem{rds}
R.~de~Sousa, Phys. Rev. B \textbf{76}, 245306 (2007)

\bibitem{fi}
L.~Faoro, L.~Ioffe, Phys. Rev. Lett. \textbf{100}, 227006 (2008)

\bibitem{longobardi2007}
L.~Longobardi, S.~Pottorf, V.~Patel, J.E. Lukens, IEEE Transactions on Applied
  Superconductivity \textbf{17}, 88 (2007)

\bibitem{pottorf2007}
S.~Pottorf, V.~Patel, J.E. Lukens, in \emph{Extended Abstracts of the 11th
  International Superconductive Electronics Conference} (2007)

\bibitem{vanharlingen2004}
D.J.V. Harlingen, T.L. Robertson, B.L.T. Plourde, P.A. Reichardt, T.A. Crane,
  J.~Clarke, Physical Review B (Condensed Matter and Materials Physics)
  \textbf{70}(6), 064517 (2004).
\newblock \doi{10.1103/PhysRevB.70.064517}.
\newblock \urlprefix\url{http://link.aps.org/abstract/PRB/v70/e064517}

\bibitem{chen2008}
W.~Chen, D.A. Bennett, V.~Patel, J.E. Lukens, {SUPERCONDUCTOR SCIENCE \&
  TECHNOLOGY} \textbf{{21}}({7}), 075013 ({2008}).
\newblock \doi{{10.1088/0953-2048/21/7/075013}}

\bibitem{bartet2005}
P.~Bertet, I.~Chiorescu, G.~Burkard, K.~Semba, C.J.P.M. Harmans, D.P.
  DiVincenzo, J.E. Mooij, Physical Review Letters \textbf{95}(25), 257002
  (2005).
\newblock \doi{10.1103/PhysRevLett.95.257002}.
\newblock \urlprefix\url{http://link.aps.org/abstract/PRL/v95/e257002}

\end{thebibliography}

\end{document}